
%
%
%
\def\unredoffs{} \def\redoffs{\voffset=-.31truein\hoffset=-.59truein}
\def\speclscape{\special{ps: landscape}}
%
%
%
%
\newbox\leftpage \newdimen\fullhsize \newdimen\hstitle \newdimen\hsbody
\tolerance=1000\hfuzz=2pt
\catcode`\@=11 
\def\bigans{b }
\def\answ{b }

%
\ifx\answ\bigans\message{(This will come out unreduced.}
\magnification=1200\unredoffs\baselineskip=16pt plus 2pt minus 1pt
\hsbody=\hsize \hstitle=\hsize 
\else\message{(This will be reduced.} \let\l@r=L
\magnification=1000\baselineskip=16pt plus 2pt minus 1pt \vsize=7truein
\redoffs \hstitle=8truein\hsbody=4.75truein\fullhsize=10truein\hsize=\hsbody
\output={\ifnum\pageno=0 
  \shipout\vbox{\speclscape{\hsize\fullhsize\makeheadline}
    \hbox to \fullhsize{\hfill\pagebody\hfill}}\advancepageno
  \else
  \almostshipout{\leftline{\vbox{\pagebody\makefootline}}}\advancepageno
  \fi}
\def\almostshipout#1{\if L\l@r \count1=1 \message{[\the\count0.\the\count1]}
      \global\setbox\leftpage=#1 \global\let\l@r=R
 \else \count1=2
  \shipout\vbox{\speclscape{\hsize\fullhsize\makeheadline}
      \hbox to\fullhsize{\box\leftpage\hfil#1}}  \global\let\l@r=L\fi}
\fi
%
\newcount\yearltd\yearltd=\year\advance\yearltd by -1900

\def\Title#1#2{\nopagenumbers\abstractfont\hsize=\hstitle\rightline{#1}%
\vskip 1in\centerline{\titlefont #2}\abstractfont\vskip .5in\pageno=0}
\def\Date#1{\vfill\leftline{#1}\tenpoint\supereject\global\hsize=\hsbody%
\footline={\hss\tenrm\folio\hss}}
%

\def\draftmode{\message{ DRAFTMODE }\def\draftdate{{\rm preliminary draft:
\number\month/\number\day/\number\yearltd\ \ \hourmin}}%
\headline={\hfil\draftdate}\writelabels\baselineskip=20pt plus 2pt minus 2pt
 {\count255=\time\divide\count255 by 60 \xdef\hourmin{\number\count255}
  \multiply\count255 by-60\advance\count255 by\time
  \xdef\hourmin{\hourmin:\ifnum\count255<10 0\fi\the\count255}}}
\def\nolabels{\def\wrlabeL##1{}\def\eqlabeL##1{}\def\reflabeL##1{}}
\def\writelabels{\def\wrlabeL##1{\leavevmode\vadjust{\rlap{\smash%
{\line{{\escapechar=` \hfill\rlap{\sevenrm\hskip.03in\string##1}}}}}}}%
\def\eqlabeL##1{{\escapechar-1\rlap{\sevenrm\hskip.05in\string##1}}}%
\def\reflabeL##1{\noexpand\llap{\noexpand\sevenrm\string\string\string##1}}}
\nolabels
%
\global\newcount\secno \global\secno=0
\global\newcount\meqno \global\meqno=1
\def\newsec#1{\global\advance\secno by1\message{(\the\secno. #1)}
\global\subsecno=0\eqnres@t\noindent{\bf\the\secno. #1}
\writetoca{{\secsym} {#1}}\par\nobreak\medskip\nobreak}
\def\eqnres@t{\xdef\secsym{\the\secno.}\global\meqno=1\bigbreak\bigskip}
\def\sequentialequations{\def\eqnres@t{\bigbreak}}\xdef\secsym{}
\global\newcount\subsecno \global\subsecno=0
\def\subsec#1{\global\advance\subsecno by1\message{(\secsym\the\subsecno. #1)}
\ifnum\lastpenalty>9000\else\bigbreak\fi
\noindent{\it\secsym\the\subsecno. #1}\writetoca{\string\quad
{\secsym\the\subsecno.} {#1}}\par\nobreak\medskip\nobreak}
\def\appendix#1#2{\global\meqno=1\global\subsecno=0\xdef\secsym{\hbox{#1.}}
\bigbreak\bigskip\noindent{\bf Appendix #1. #2}\message{(#1. #2)}
\writetoca{Appendix {#1.} {#2}}\par\nobreak\medskip\nobreak}
%
%
\def\eqnn#1{\xdef #1{(\secsym\the\meqno)}\writedef{#1\leftbracket#1}%
\global\advance\meqno by1\wrlabeL#1}
\def\eqna#1{\xdef #1##1{\hbox{$(\secsym\the\meqno##1)$}}
\writedef{#1\numbersign1\leftbracket#1{\numbersign1}}%
\global\advance\meqno by1\wrlabeL{#1$\{\}$}}
\def\eqn#1#2{\xdef #1{(\secsym\the\meqno)}\writedef{#1\leftbracket#1}%
\global\advance\meqno by1$$#2\eqno#1\eqlabeL#1$$}
%
\newskip\footskip\footskip14pt plus 1pt minus 1pt 
\def\footnotefont{\ninepoint}\def\f@t#1{\footnotefont #1\@foot}
\def\f@@t{\baselineskip\footskip\bgroup\footnotefont\aftergroup\@foot\let\next}
\setbox\strutbox=\hbox{\vrule height9.5pt depth4.5pt width0pt}
\global\newcount\ftno \global\ftno=0
\def\foot{\global\advance\ftno by1\footnote{$^{\the\ftno}$}}
%
\newwrite\ftfile
\def\footend{\def\foot{\global\advance\ftno by1\chardef\wfile=\ftfile
$^{\the\ftno}$\ifnum\ftno=1\immediate\openout\ftfile=foots.tmp\fi%
\immediate\write\ftfile{\noexpand\smallskip%
\noexpand\item{f\the\ftno:\ }\pctsign}\findarg}%
\def\footatend{\vfill\eject\immediate\closeout\ftfile{\parindent=20pt
\centerline{\bf Footnotes}\nobreak\bigskip\input foots.tmp }}}
\def\footatend{}
%
%
\global\newcount\refno \global\refno=1
\newwrite\rfile
\def\ref{[\the\refno]\nref}
\def\nref#1{\xdef#1{[\the\refno]}\writedef{#1\leftbracket#1}%
\ifnum\refno=1\immediate\openout\rfile=refs.tmp\fi
\global\advance\refno by1\chardef\wfile=\rfile\immediate
\write\rfile{\noexpand\item{#1\ }\reflabeL{#1\hskip.31in}\pctsign}\findarg}
\def\findarg#1#{\begingroup\obeylines\newlinechar=`\^^M\pass@rg}
{\obeylines\gdef\pass@rg#1{\writ@line\relax #1^^M\hbox{}^^M}%
\gdef\writ@line#1^^M{\expandafter\toks0\expandafter{\striprel@x #1}%
\edef\next{\the\toks0}\ifx\next\em@rk\let\next=\endgroup\else\ifx\next\empty%
\else\immediate\write\wfile{\the\toks0}\fi\let\next=\writ@line\fi\next\relax}}
\def\striprel@x#1{} \def\em@rk{\hbox{}}
\def\lref{\begingroup\obeylines\lr@f}
\def\lr@f#1#2{\gdef#1{\ref#1{#2}}\endgroup\unskip}

\def\addref#1{\immediate\write\rfile{\noexpand\item{}#1}} 
\def\startrefs#1{\immediate\openout\rfile=refs.tmp\refno=#1}
\def\xref{\expandafter\xr@f}\def\xr@f[#1]{#1}
\def\refs#1{\count255=1[\r@fs #1{\hbox{}}]}
\def\r@fs#1{\ifx\und@fined#1\message{reflabel \string#1 is undefined.}%
\nref#1{need to supply reference \string#1.}\fi%
\vphantom{\hphantom{#1}}\edef\next{#1}\ifx\next\em@rk\def\next{}%
\else\ifx\next#1\ifodd\count255\relax\xref#1\count255=0\fi%
\else#1\count255=1\fi\let\next=\r@fs\fi\next}
%

%
\newwrite\ffile\global\newcount\figno \global\figno=1
\def\fig{fig.~\the\figno\nfig}
\def\nfig#1{\xdef#1{fig.~\the\figno}%
\writedef{#1\leftbracket fig.\noexpand~\the\figno}%
\ifnum\figno=1\immediate\openout\ffile=figs.tmp\fi\chardef\wfile=\ffile%
\immediate\write\ffile{\noexpand\medskip\noexpand\item{Fig.\ \the\figno. }
\reflabeL{#1\hskip.55in}\pctsign}\global\advance\figno by1\findarg}
\def\xfig{\expandafter\xf@g}\def\xf@g fig.\penalty\@M\ {}
\def\figs#1{figs.~\f@gs #1{\hbox{}}}
\def\f@gs#1{\edef\next{#1}\ifx\next\em@rk\def\next{}\else
\ifx\next#1\xfig #1\else#1\fi\let\next=\f@gs\fi\next}
\newwrite\lfile
{\escapechar-1\xdef\pctsign{\string\%}\xdef\leftbracket{\string\{}
\xdef\rightbracket{\string\}}\xdef\numbersign{\string\#}}

\def\writestop{\def\writestoppt{\immediate\write\lfile{\string\pageno%
\the\pageno\string\startrefs\leftbracket\the\refno\rightbracket%
\string\def\string\secsym\leftbracket\secsym\rightbracket%
\string\secno\the\secno\string\meqno\the\meqno}\immediate\closeout\lfile}}
\def\writestoppt{}\def\writedef#1{}
\def\seclab#1{\xdef #1{\the\secno}\writedef{#1\leftbracket#1}\wrlabeL{#1=#1}}
\def\subseclab#1{\xdef #1{\secsym\the\subsecno}%
\writedef{#1\leftbracket#1}\wrlabeL{#1=#1}}
\newwrite\tfile \def\writetoca#1{}
\def\leaderfill{\leaders\hbox to 1em{\hss.\hss}\hfill}
\def\writetoc{\immediate\openout\tfile=toc.tmp
   \def\writetoca##1{{\edef\next{\write\tfile{\noindent ##1
   \string\leaderfill {\noexpand\number\pageno} \par}}\next}}}

%
\catcode`\@=12 
%
\edef\tfontsize{\ifx\answ\bigans scaled\magstep3\else scaled\magstep4\fi}
\font\titlerm=cmr10 \tfontsize \font\titlerms=cmr7 \tfontsize
\font\titlermss=cmr5 \tfontsize \font\titlei=cmmi10 \tfontsize
\font\titleis=cmmi7 \tfontsize \font\titleiss=cmmi5 \tfontsize
\font\titlesy=cmsy10 \tfontsize \font\titlesys=cmsy7 \tfontsize
\font\titlesyss=cmsy5 \tfontsize \font\titleit=cmti10 \tfontsize
\skewchar\titlei='177 \skewchar\titleis='177 \skewchar\titleiss='177
\skewchar\titlesy='60 \skewchar\titlesys='60 \skewchar\titlesyss='60
\def\titlefont{\def\rm{\fam0\titlerm}
\textfont0=\titlerm \scriptfont0=\titlerms \scriptscriptfont0=\titlermss
\textfont1=\titlei \scriptfont1=\titleis \scriptscriptfont1=\titleiss
\textfont2=\titlesy \scriptfont2=\titlesys \scriptscriptfont2=\titlesyss
\textfont\itfam=\titleit \def\it{\fam\itfam\titleit}\rm}
 \ifx\answ\bigans\else scaled\magstep1\fi
\ifx\answ\bigans\def\abstractfont{\tenpoint}\else
\font\abssl=cmsl10 scaled \magstep1
\font\absrm=cmr10 scaled\magstep1 \font\absrms=cmr7 scaled\magstep1
\font\absrmss=cmr5 scaled\magstep1 \font\absi=cmmi10 scaled\magstep1
\font\absis=cmmi7 scaled\magstep1 \font\absiss=cmmi5 scaled\magstep1
\font\abssy=cmsy10 scaled\magstep1 \font\abssys=cmsy7 scaled\magstep1
\font\abssyss=cmsy5 scaled\magstep1 \font\absbf=cmbx10 scaled\magstep1
\skewchar\absi='177 \skewchar\absis='177 \skewchar\absiss='177
\skewchar\abssy='60 \skewchar\abssys='60 \skewchar\abssyss='60
\def\abstractfont{\def\rm{\fam0\absrm}
\textfont0=\absrm \scriptfont0=\absrms \scriptscriptfont0=\absrmss
\textfont1=\absi \scriptfont1=\absis \scriptscriptfont1=\absiss
\textfont2=\abssy \scriptfont2=\abssys \scriptscriptfont2=\abssyss
\textfont\itfam=\bigit \def\it{\fam\itfam\bigit}\def\footnotefont{\tenpoint}%
\textfont\slfam=\abssl \def\sl{\fam\slfam\abssl}%
\textfont\bffam=\absbf \def\bf{\fam\bffam\absbf}\rm}\fi
\def\tenpoint{\def\rm{\fam0\tenrm}
\textfont0=\tenrm \scriptfont0=\sevenrm \scriptscriptfont0=\fiverm
\textfont1=\teni  \scriptfont1=\seveni  \scriptscriptfont1=\fivei
\textfont2=\tensy \scriptfont2=\sevensy \scriptscriptfont2=\fivesy
\textfont\itfam=\tenit \def\it{\fam\itfam\tenit}\def\footnotefont{\ninepoint}%
\textfont\bffam=\tenbf \def\bf{\fam\bffam\tenbf}\def\sl{\fam\slfam\tensl}\rm}
\font\ninerm=cmr9 \font\sixrm=cmr6 \font\ninei=cmmi9 \font\sixi=cmmi6
\font\ninesy=cmsy9 \font\sixsy=cmsy6 \font\ninebf=cmbx9
\font\nineit=cmti9 \font\ninesl=cmsl9 \skewchar\ninei='177
\skewchar\sixi='177 \skewchar\ninesy='60 \skewchar\sixsy='60
\def\ninepoint{\def\rm{\fam0\ninerm}
\textfont0=\ninerm \scriptfont0=\sixrm \scriptscriptfont0=\fiverm
\textfont1=\ninei \scriptfont1=\sixi \scriptscriptfont1=\fivei
\textfont2=\ninesy \scriptfont2=\sixsy \scriptscriptfont2=\fivesy
\textfont\itfam=\ninei \def\it{\fam\itfam\nineit}\def\sl{\fam\slfam\ninesl}%
\textfont\bffam=\ninebf \def\bf{\fam\bffam\ninebf}\rm}
%
%

\hyphenation{anom-aly anom-alies coun-ter-term coun-ter-terms}
\def\inv{^{\raise.15ex\hbox{${\scriptscriptstyle -}$}\kern-.05em 1}}

\def\Dsl{\,\raise.15ex\hbox{/}\mkern-13.5mu D} 
\def\dsl{\raise.15ex\hbox{/}\kern-.57em\partial}

\font\bigit=cmti10 scaled \magstep1
\def\lspace{\ifx\answ\bigans{}\else\qquad\fi}
\def\lbspace{\ifx\answ\bigans{}\else\hskip-.2in\fi} 
\def\boxeqn#1{\vcenter{\vbox{\hrule\hbox{\vrule\kern3pt\vbox{\kern3pt
    \hbox{${\displaystyle #1}$}\kern3pt}\kern3pt\vrule}\hrule}}}
\def\mbox#1#2{\vcenter{\hrule \hbox{\vrule height#2in
        \kern#1in \vrule} \hrule}}  
%

\def\darr#1{\raise1.5ex\hbox{$\leftrightarrow$}\mkern-16.5mu #1}

\def\roughly#1{\raise.3ex\hbox{$#1$\kern-.75em\lower1ex\hbox{$\sim$}}}

\let\includefigures=\iftrue
\let\useblackboard=\iftrue
\newfam\black

\includefigures
\message{If you do not have epsf.tex (to include figures),}
\message{change the option at the top of the tex file.}
\input epsf
\def\figin{\epsfcheck\figin}\def\figins{\epsfcheck\figins}
\def\epsfcheck{\ifx\epsfbox\UnDeFiNeD
\message{(NO epsf.tex, FIGURES WILL BE IGNORED)}
\gdef\figin##1{\vskip2in}\gdef\figins##1{\hskip.5in}
\else\message{(FIGURES WILL BE INCLUDED)}%
\gdef\figin##1{##1}\gdef\figins##1{##1}\fi}
\def\DefWarn#1{}
\def\figinsert{\goodbreak\midinsert}
\def\ifig#1#2#3{\DefWarn#1\xdef#1{fig.~\the\figno}
\writedef{#1\leftbracket fig.\noexpand~\the\figno}%
\figinsert\figin{\centerline{#3}}\medskip\centerline{\vbox{
\baselineskip12pt\advance\hsize by -1truein
\noindent\footnotefont{\bf Fig.~\the\figno:} #2}}
\endinsert\global\advance\figno by1}
\else
\def\ifig#1#2#3{\xdef#1{fig.~\the\figno}
\writedef{#1\leftbracket fig.\noexpand~\the\figno}%
\global\advance\figno by1} \fi

\def\id{{1 \kern-.28em {\rm l}}}

\def\A{{\cal A}}

\def\K3{{\bf K3}}
\def\journal#1&#2(#3){\unskip, \sl #1\ \bf #2 \rm(19#3) }
\def\andjournal#1&#2(#3){\sl #1~\bf #2 \rm (19#3) }

\def\tilde{\widetilde}

\def\frac#1#2{{#1\over#2}}

\def\inbar{\,\vrule height1.5ex width.4pt depth0pt}
\def\IC{\relax\hbox{$\inbar\kern-.3em{\rm C}$}}
\def\IR{\relax{\rm I\kern-.18em R}}
\def\IP{\relax{\rm I\kern-.18em P}}

%
%

%
\catcode`\@=11
\def\slash#1{\mathord{\mathpalette\c@ncel{#1}}}
\overfullrule=0pt

\def\NN{{\cal N}}
\def\OO{{\cal O}}

\def\underrel#1\over#2{\mathrel{\mathop{\kern\z@#1}\limits_{#2}}}

\catcode`\@=12


%


\def\ra{{\rightarrow}}

\def\IS{{\bf S}}

\lref\wenniu{
  X.~G.~Wen and Q.~Niu,
  ``Ground State Degeneracy Of The FQH States In Presence
  Of Random Potential And On High Genus Riemann Surfaces,'', Phys. \ Rev.  {\bf B41}, 9377~(1990)
}

\lref\WenQN{
  X.~G.~Wen,
  ``Topological orders and edge excitations in FQH states,''
Advances in Physics, {\bf 44}, 405 (1995).
}

\lref\KitaevDM{
  A.~Kitaev and J.~Preskill,
  ``Topological entanglement entropy,''
  Phys.\ Rev.\ Lett.\  {\bf 96}, 110404 (2006)
  [arXiv:hep-th/0510092].
}

\lref\levin{ M. Levin, X.-G. Wen,  ''Detecting topological order in a ground state wave function,''
             Phys. Rev. Lett., {\bf 96}, 110405 (2006).
}

\lref\SchwimmerYH{
  A.~Schwimmer and S.~Theisen,
  ``Entanglement Entropy, Trace Anomalies and Holography,''
  arXiv:0802.1017 [hep-th].
}

\lref\SolodukhinDH{
  S.~N.~Solodukhin,
  ``Entanglement entropy, conformal invariance and extrinsic geometry,''
  arXiv:0802.3117 [hep-th].
}

\lref\WittenZW{
  E.~Witten,
  ``Anti-de Sitter space, thermal phase transition, and confinement in  gauge
  theories,''
  Adv.\ Theor.\ Math.\ Phys.\  {\bf 2}, 505 (1998)
  [arXiv:hep-th/9803131].
}

\lref\KlebanovWS{
  I.~R.~Klebanov, D.~Kutasov and A.~Murugan,
  ``Entanglement as a Probe of Confinement,''
  arXiv:0709.2140 [hep-th].
}

\lref\RT{
  S.~Ryu and T.~Takayanagi,
  ``Holographic derivation of entanglement entropy from AdS/CFT,''
  Phys.\ Rev.\ Lett.\  {\bf 96}, 181602 (2006)
  [arXiv:hep-th/0603001];
  ``Aspects of holographic entanglement entropy,''
  JHEP {\bf 0608}, 045 (2006)
  [arXiv:hep-th/0605073].
}

\lref\SolodukhinDH{
  S.~N.~Solodukhin,
  ``Entanglement entropy, conformal invariance and extrinsic geometry,''
  arXiv:0802.3117 [hep-th].
}

\lref\TsuiYY{
  D.~C.~Tsui, H.~L.~Stormer and A.~C.~Gossard,
  ``Two-dimensional magnetotransport in the extreme quantum limit,''
  Phys.\ Rev.\ Lett.\  {\bf 48}, 1559 (1982).
}

\lref\WenIV{
  X.~G.~Wen,
  ``Topological order in rigid states,''
  Int.\ J.\ Mod.\ Phys.\  B {\bf 4}, 239 (1990).
}

\lref\WenDeg{
  Wen, X. G.,
  ``Vacuum degeneracy of chiral spin states in compactified space,''
  Phys. \ Rev.  {\bf B40}, 7387 (1989).
}

\lref\ArovasQR{
  D.~Arovas, J.~R.~Schrieffer and F.~Wilczek,
  ``Fractional Statistics And The Quantum Hall Effect,''
  Phys.\ Rev.\ Lett.\  {\bf 53}, 722 (1984).
}

\lref\FendleyGR{
  P.~Fendley, M.~P.~A.~Fisher and C.~Nayak,
  ``Topological Entanglement Entropy from the Holographic Partition Function,''
  J.\ Statist.\ Phys.\  {\bf 126}, 1111 (2007)
  [arXiv:cond-mat/0609072].
}
\lref\DongFT{
  S.~Dong, E.~Fradkin, R.~G.~Leigh and S.~Nowling,
  ``Topological Entanglement Entropy in Chern-Simons Theories and Quantum Hall
  Fluids,''
  arXiv:0802.3231 [hep-th].
}

\lref\PeetWN{
  A.~W.~Peet and J.~Polchinski,
  ``UV/IR relations in AdS dynamics,''
  Phys.\ Rev.\  D {\bf 59}, 065011 (1999)
  [arXiv:hep-th/9809022].
}

\lref\BuividovichKQ{
  P.~V.~Buividovich and M.~I.~Polikarpov,
  ``Numerical study of entanglement entropy in SU(2) lattice gauge theory,''
  arXiv:0802.4247 [hep-lat].
}

\lref\VelytskyRS{
  A.~Velytsky,
  ``Entanglement entropy in d+1 SU(N) gauge theory,''
  arXiv:0801.4111 [hep-th].
}

\lref\HirataJX{
  T.~Hirata and T.~Takayanagi,
  ``AdS/CFT and strong subadditivity of entanglement entropy,''
  JHEP {\bf 0702}, 042 (2007)
  [arXiv:hep-th/0608213].
}

\lref\Nayak{
  C.~Nayak and F.~Wilczek,
  ``$2n$ Quasihole States Realize $2^{n-1}$-Dimensional Spinor Braiding Statistics in Paired Quantum Hall States,''
  Nucl. Phys. {\bf B 479}, 529 (1996)
  [arXiv:cond-mat/9605145].
Read, N.  and Rezayi, E., ''Quasiholes and fermionic zero modes of paired fractional quantum Hall states: The mechanism for non-Abelian statistics,''
Phys. Rev. {\bf B 54} 16864 [arXiv:cond-mat/9609079].
}

\lref\HeadrickKM{
  M.~Headrick and T.~Takayanagi,
  ``A holographic proof of the strong subadditivity of entanglement entropy,''
  Phys.\ Rev.\  D {\bf 76}, 106013 (2007)
  [arXiv:0704.3719 [hep-th]].
}

\lref\WittenZW{
  E.~Witten,
  ``Anti-de Sitter space, thermal phase transition, and confinement in  gauge
  theories,''
  Adv.\ Theor.\ Math.\ Phys.\  {\bf 2}, 505 (1998)
  [arXiv:hep-th/9803131].
}

\lref\HubenyXT{
  V.~E.~Hubeny, M.~Rangamani and T.~Takayanagi,
  ``A covariant holographic entanglement entropy proposal,''
  JHEP {\bf 0707}, 062 (2007)
  [arXiv:0705.0016 [hep-th]].
}
\lref\HammaUD{
  A.~Hamma, P.~Zanardi and X.~G.~Wen,
  ``String and membrane condensation on 3D lattices,''
  Phys.\ Rev.\  B {\bf 72}, 035307 (2005)
  [arXiv:cond-mat/0411752].
 H.~Bombin and M.~A.~Martin-Delgado,
  ``Exact Topological Quantum Order in D=3 and Beyond: Branyons and Brane-Net Condensates,''
  Phys.\ Rev.\  B {\bf 75}, 075103 (2007)
  [arXiv:cond-mat/0607736].
}

\lref\BombelliRW{
  L.~Bombelli, R.~K.~Koul, J.~H.~Lee and R.~D.~Sorkin,
  ``A Quantum Source of Entropy for Black Holes,''
  Phys.\ Rev.\  D {\bf 34}, 373 (1986).
}
\lref\SrednickiIM{
  M.~Srednicki,
  ``Entropy and area,''
  Phys.\ Rev.\ Lett.\  {\bf 71}, 666 (1993)
  [arXiv:hep-th/9303048].
}

\lref\MaldacenaRE{
  J.~M.~Maldacena,
  ``The large N limit of superconformal field theories and supergravity,''
  Adv.\ Theor.\ Math.\ Phys.\  {\bf 2}, 231 (1998)
  [Int.\ J.\ Theor.\ Phys.\  {\bf 38}, 1113 (1999)]
  [arXiv:hep-th/9711200].
  S.~S.~Gubser, I.~R.~Klebanov and A.~M.~Polyakov,
  ``Gauge theory correlators from non-critical string theory,''
  Phys.\ Lett.\  B {\bf 428}, 105 (1998)
  [arXiv:hep-th/9802109].
 E.~Witten,
  ``Anti-de Sitter space and holography,''
  Adv.\ Theor.\ Math.\ Phys.\  {\bf 2}, 253 (1998)
  [arXiv:hep-th/9802150].
}

\lref\BrodieYZ{
  J.~H.~Brodie, L.~Susskind and N.~Toumbas,
  ``How Bob Laughlin tamed the giant graviton from Taub-NUT space,''
  JHEP {\bf 0102}, 003 (2001)
  [arXiv:hep-th/0010105].
  S.~S.~Gubser and M.~Rangamani,
  ``D-brane dynamics and the quantum Hall effect,''
  JHEP {\bf 0105}, 041 (2001)
  [arXiv:hep-th/0012155].
}

\lref\BarbonUT{
  J.~L.~F.~Barbon and C.~A.~Fuertes,
  ``Holographic entanglement entropy probes (non)locality,''
  JHEP {\bf 0804}, 096 (2008)
  [arXiv:0803.1928 [hep-th]].
}

\lref\MateosNU{
  D.~Mateos, R.~C.~Myers and R.~M.~Thomson,
  ``Holographic phase transitions with fundamental matter,''
  Phys.\ Rev.\ Lett.\  {\bf 97}, 091601 (2006)
  [arXiv:hep-th/0605046].
}

\lref\FursaevIH{
  D.~V.~Fursaev,
  ``Proof of the holographic formula for entanglement entropy,''
  JHEP {\bf 0609}, 018 (2006)
  [arXiv:hep-th/0606184].
}

\lref\NishiokaGR{
  T.~Nishioka and T.~Takayanagi,
  ``AdS bubbles, entropy and closed string tachyons,''
  JHEP {\bf 0701}, 090 (2007)
  [arXiv:hep-th/0611035].
}

\lref\BarbonSR{
  J.~L.~F.~Barbon and C.~A.~Fuertes,
  ``A note on the extensivity of the holographic entanglement entropy,''
  JHEP {\bf 0805}, 053 (2008)
  [arXiv:0801.2153 [hep-th]].
}

\Title{\vbox{\baselineskip12pt
\hbox{YITP-SB-08-19}
}}
{\vbox{\centerline{Topological Entanglement Entropy and Holography}
\vskip.06in
}}
\centerline{Ari Pakman and Andrei Parnachev}
\bigskip
\centerline{{\it C.N.Yang Institute for Theoretical Physics }}
\centerline{\it Stony Brook University }
\centerline{\it Stony Brook, NY 11794-3840, USA}
\vskip.1in \vskip.1in \centerline{\bf Abstract}
\noindent
We study the entanglement entropy in confining theories with gravity
duals
using  the holographic prescription of Ryu and Takayanagi.
The entanglement
entropy between a region and its complement is proportional to
the minimal area of a bulk hypersurface ending
on their border.
We consider a disk in 2+1 dimensions and a ball in 3+1 dimensions
and find in both cases two types of bulk hypersurfaces with
different topology, similar to the case of the slab geometry
considered by Klebanov, Kutasov and Murugan. Depending on the value
of the radius, one or the other type of hypersurfaces dominates the
calculation of entanglement entropy.
In 2+1 dimensions a useful measure of topological order of the ground state
is the topological entanglement entropy, which is defined to be the constant term
in the entanglement entropy of a disk in the limit of large radius.
We compute this quantity and find that it vanishes for confining gauge theory,
in accord with our expectations.
%
In 3+1 dimensions the analogous quantity is shown to be generically nonzero
and cutoff-dependent.

\vfill

\Date{May 2008}



\newsec{Introduction and summary}

Certain systems in two spacial dimensions are known to exhibit topological order~\refs{\wenniu,\WenQN}.
In this case the ground state of the system has highly non-trivial properties,
related to the degeneracy on  higher genus surfaces and the existence
of quasi-particle excitations.
Such a non-local order can exist even when  local order parameters
vanish.
Recently a quantity called ``topological entanglement entropy'' was proposed
\refs{\KitaevDM,\levin} to measure the degree of topological entanglement.
It involves computing the entanglement entropy of a large disk and extracting the constant
term, which does not scale with the circumference of the disk.
The natural generalization to higher dimensions would involve entanglement
entropy of a large ball.

Computing the entanglement entropy for a general system is not a trivial
exercise.
Recently Ryu and Takayanagi \RT\ proposed a way to compute the entanglement entropy
in theories with holographic dual.
The computation involves finding a certain minimal surface in the bulk of
the asymptotically AdS space.\foot{
Schwimmer and Theisen \SchwimmerYH\ argued that anomalous terms in the
entanglement entropy are not correctly reproduced in the holographic prescription.
However the validity of this claim has been questioned in \SolodukhinDH.
The strong subadditivity  of the holographic entanglement entropy has been
studied in  \refs{\HirataJX,\HeadrickKM}. Covariant proposal for
computing entanglement entropy was formulated in~\HubenyXT.
In \BarbonUT\ entanglement entropy in Little String Theory has been studied.
Other interesting work includes \refs{\FursaevIH\NishiokaGR-\BarbonSR}. }

This hypersurface approaches the border between the region and its
complement on the boundary.
The generalization of this prescription also exists for
the theories whose duals are not asymptotically AdS.
In this paper we make use of this proposal to compute the topological entanglement
entropy of some simple confining theories in 2+1 and 3+1 dimensions.

More precisely, we will study D3 and D4 branes compactified on a circle, which
at large distances become confining 2+1 and 3+1 dimensional theories respectively.
The entanglement entropy of the ``slab'' (the subspace defined by $-\ell/2<x<\ell/2$, where
$x$ is one of the spacial coordinates) in these systems has been studied in \KlebanovWS.
As it turns out, there are two types of minimal surfaces: connected and disconnected ones.
In ref. \KlebanovWS\ a first order phase transition between these two types
of solutions was observed as a function of $\ell$.
For small $\ell$ the connected solution dominates the computation
of entanglement entropy, while for large $\ell$
the disconnected solution becomes preferred.
The authors interpreted this phase transition as a signature of confinement.
In the case of a disk in 2+1 dimensions or a ball in 3+1 dimensions
there is again a single parameter $R$, and
as we will see below there are again two types of solutions in the bulk.
The analog of a connected solution has a disk topology and dominates at smaller
values of $R$.
In this solution the circle (sphere) which is the boundary of the disk (ball)
shrinks to zero size in the bulk, while Kaluza-Klein circle remains finite.
The analog of the disconnected solution has a cylinder topology and
dominates at larger values of $R$.
This is the solution where the Kaluza-Klein circle shrinks to
zero size.
For intermediate values of $R$ the structure has a classic ``swallowtail''
shape typical of first order phase transitions.

To compute the topological entanglement entropy in 2+1 dimensions,
we need to consider a disk whose radius is very large compared
to the correlation length.
In this regime the only available solution has the topology of
a cylinder.
We show that this solution approaches the straight cylinder as $R\ra\infty$,
and the topological entropy associated to it vanishes.
In hindsight, this is not very surprising, since we do not expect long-range
topological order in the ground state of $2+1$ dimensional QCD.
Our calculation hence can be viewed as a consistency check on both
the holographic prescription for computing entanglement entropy and
topological entanglement entropy.
Note that the existence of phase transition between the two topologies
is crucial for the whole picture to be consistent.

We then compute the entanglement entropy in the theory on $D4$ branes compactified
on a circle (this theory has many features of QCD in 3+1 dimensions, as emphasized in \WittenZW).
The structure of the solutions is similar to the 2+1 dimensional case.
There is again a first order phase transition between the hypersurfaces of disk
and cylinder topology, as $R$ is varied.
However when computing the radius-independent term we encounter a
surprise: it contains an explicit dependence on the cutoff.
We attribute this to the unconventional UV properties of the theory
(see discussion Section).

The rest of the paper is organized as follows.
In the next Section we review the definitions of entanglement and topological
entanglement entropies and the holographic proposal of Ryu and Takayanagi.
In Section 3 we start with a warm-up example of a cylinder in $\NN=4$ super
Yang Mills in 3+1 dimensions and investigate the structure of
the solutions.
We then compactify the theory on a circle to generate a confining theory in
$2+1$ dimensions.
We re-investigate the structure of the solutions, compute the value
of entanglement entropy as a function of $R$ and extract the value of
topological entanglement entropy, which vanishes.
In Section 4 we repeat this analysis for the case of the
$D4$ branes compactified on a circle.
We find a very similar phase transition structure, while the
generalization of topological entanglement entropy turns out to be cutoff-dependent.
We discuss our results in Section 4.
Appendix contains some details regarding the structure of the
minimal hypersurfaces.

\newsec{Review of the background material}

\subsec{Entanglement entropy and the holographic prescription}
Consider a pure quantum state $|\Psi \rangle$ in a system that can be subdivided into two subsystems $A$ and $B$.
If we trace  its density matrix over the degrees of freedom of  $B$,
\eqn\densb{
\rho_A = Tr_B \left(| \Psi \rangle \langle \Psi | \right) \,,
}
the density matrix $\rho_A$ will be in general in a mixed state.
Its  von Neumann entropy
\eqn\enta{
S_A = -Tr_A \left( \rho_A \ln \rho_A \right)
}
is called the {\it entanglement entropy} and is a measure of the entanglement between  $A$ and $B$ in the original state~$|\Psi \rangle$.
In particular, it is zero if $|\Psi \rangle$ is the product of a state in $A$ and a state in $B$, and one can
verify that $S_A=S_B$.

Of particular interest is the case when the two subsystems $A$ and $B$ correspond
to two regions in the spacetime of a local quantum field
theory and $|\Psi\rangle$ is the ground state \refs{\BombelliRW,\SrednickiIM}.
In this case the entanglement entropy is cutoff dependent
and its leading term is proportional to the area of the surface that separates the two regions.

For quantum field theories which have gravitational duals \MaldacenaRE, the authors of \RT\  proposed a simple
geometric prescription for computing the entanglement entropy of the vacuum.
For a  $d+1$ dimensional conformal field theory with an $AdS_{d+2}$ dual,
the idea is to find a $d$-dimensional surface~$\Gamma$ which minimizes the action
\eqn\entactionc{
S_A = \frac{1}{4G_N^{(d+2)}} \int_{\Gamma} d^d \sigma  \sqrt{G_{ind}^{(d)}}
}
and approaches the boundary of the regions $A$ and $B$ at the boundary of the $AdS_{d+2}$ manifold.
The surface $\Gamma$ is defined at a fixed time and   $G_{ind}^{(d)}$ is the induced string frame metric.
As shown in \RT, for the case of $AdS_3$, this prescription reproduces precisely the known expression
for the entaglement entropy in 2D CFT, and has been subjected to several tests.

When the boundary theory is not conformal, which is the case we will be interested in this work,
the above functional generalizes to
\eqn\entaction{
S_A = \frac{1}{4G_N^{(10)}} \int_{\Gamma} d^8 \sigma e^{- 2\phi} \sqrt{G_{ind}^{(8)}}
}
where now  $\Gamma$  extends  in all  remaining directions.

\subsec{Measures of quantum order and topological entanglement entropy}
Following the discovery of the Fractional Quantum Hall  states \TsuiYY,  it  became clear that there
are quantum  systems in which different phases or orders  ({\it e.g.} quantum Hall states with different filling fractions) preserve the same symmetries.
At zero temperature, different orders have different quantum correlations between the microscopic degrees
of freedom, but cannot be distinguished by any local order parameter.
For these {\it quantum orders}, Landau's theory of symmetry-breaking and local order parameter
is inadequate, and novel quantum numbers are needed to characterize them.

When the system  has an energy gap the quantum order is called {\it topological order} \WenIV,
and the theory is described in the infrared by a topological quantum field theory.
Useful characterizations of topological orders are   the degeneracy (in the infinite volume limit)
of the ground state in a Riemann surface as a function of the genus~\refs{\WenDeg,\wenniu}, the spectrum of quasiparticle excitations~\ArovasQR\
and  the structure of the gapless edge excitations~\WenQN.

Another quantum number which characterizes   topological order is the total quantum dimension, defined as follows.
The number of linearly-independent states having~$N$ quasiparticles of type $a$,
for large $N$, is proportional to  $d_a^N$, where $d_a$ is the quantum dimension of the quasiparticle $a$ \Nayak.
The {\it total quantum dimension} of the system is
\eqn\quantdim{
{\cal D} = \sqrt{\sum_a d_a^2} \,.
}
In general,  ${\cal D}>1$ signals a topological order.
In theories with no quasiparticle excitations, as in the model we consider in the next section, only the
identity sector contributes to~\quantdim, so  we expect ${\cal D}= d_{I}=1$.

As it turns out, the total quantum dimension ${\cal D}$ is intimately related to the entanglement properties of the ground state.
Consider, in $2+1$ dimensions, a disk $A$ with a smooth boundary of length~$L$ in the infinite  plane.
It was shown in~\refs{\KitaevDM,\levin} that the entanglement entropy of the ground state between the disk $A$ and its exterior
behaves as
\eqn\entent{
S_A = \alpha L -\log {\cal D} + \cdots
}
where $\alpha$ is non-universal and cutoff dependent and the additional terms vanish in the limit~$L \rightarrow \infty$.
The quantity
\eqn\topent{
\gamma = \log {\cal D}
}
is called {\it topological entanglement entropy}.
It is a measure of topological order encoded in the wave-function,
and not in the spectral properties of the Hamiltonian, as the other characterizations of topological order
mentioned above. For theories in $2+1$ dimensions whose boundary degrees of freedom live in a 2D conformal field theory,
the topological entanglement entropy can be expressed in terms of the modular matrix of the latter. For studies
exploiting this idea see \refs{\FendleyGR, \DongFT}.

The definition of $\gamma$ can be naturally generalized to dimensions higher than $2+1$ by considering
a ball instead of a disk, but these cases have not been thoroughly studied  yet.\foot{Solvable systems
exhibiting topological order in $3+1$ dimensions were studied in~\HammaUD.}

\newsec{D3 branes on a circle}

\noindent Here we consider entanglement entropy for a disk
in the strongly coupled four-dimensional $\NN=4$ SYM
compactified on a circle.
At low energies this theory reduces to the three-dimensional
gauge theory with confinement, mass gap and finite correlation length.
The metric can be written as
\eqn\metrica{  ds^2= \left({U\over L}\right)^2 \left[ \left({L\over U}\right)^4 {dU^2\over h(U)}+dx_\mu dx^\mu+h(U) dx_3^2\right]
                   +L^2 d\Omega_5^2,\qquad \mu=0,1,2    }
where
\eqn\hdef{  h(U) = 1-\left({U_0\over U}\right)^4,\qquad U_0^2 = {L^4\over 4 R_3^2} }
and $R_3$ is the radius of the Kaluza-Klein circle, and $L^4=4\pi\lambda=4\pi g_s N_c$
sets the curvature scale of the AdS space.\foot{Here and in the rest of the paper we set $\alpha'=1$.}
It is convenient to switch to the variable $z=L^2/U$.
The  metric \metrica\ takes the following form
\eqn\metricb{  ds^2= L^2 \left[ {dz^2\over z^2 h(z)} +{dx_\mu dx^\mu\over z^2} +h(z) {dx_3^2\over z^2} +d\Omega_5^2\right] }
where
\eqn\hdefa{  h(z)= 1-{z^4\over z_0^4},\qquad z_0=2 R_3  }
We follow \RT, where the prescription to compute entanglement entropy
was formulated in the holographic setup.
Consider a disk in the $x_1-x_2$ plane bounded by a circle of radius $R$,
denoted by $\IS^1_\phi$ below.
The subscript here refers to the angular coordinate in the $x_1-x_2$ plane;
the radial coordinate is denoted by $r$.
As reviewed in the previous Section, the entanglement entropy for this disk can be computed
by evaluating \entactionc\
on the minimal 8-dimensional surface at $t=const$ which asymptotes
to $\IS^1_\phi \times \IS^1_{x^3} \times \IS^5$ on the boundary at $z=0$.
In  this computation we need to set $G_N^{(10)}=8\pi^6 g_s^2$ .
The minimal surface is determined by specifying $z(r)$, with the induced metric $G^{(8)}_{\rm ind}$ given by
\eqn\indm{  ds^2_{\rm ind} =  L^2 \left[ {1\over z^2}\left( 1+{(z')^2\over h(z)}\right) dr^2  +{r^2\over z^2} d\phi^2 +
             {h(z)\over z^2} dx_3^2 + d\Omega_5^2\right] }
%
%

%
%

\subsec{Conformal limit}
\noindent We start with the warm-up example, which corresponds to the conformal theory,
where the bulk geometry is  $AdS_5\times S^5$.
In this limit $R_3\ra\infty$, $h(z)=1$ and the region in the 3+1 dimensional boundary
theory is the cylinder whose length we denote by $l$
(of course, in the compactified case, $l=2\pi R_3$).
Eq. \entactionc\ can be written as
\eqn\actconf{  S={4 N_c^2 l\over 15\pi} \int dr {r\over z^3} \sqrt{ 1+(z')^2 }}
which gives the following equation of motion for the hypersurface:
\eqn\eomconf{  {d\over dr} \left( r z'\over z^3\sqrt{1+(z')^2}\right) =
              -{3 r \sqrt{1+(z')^2}\over z^4} }
One needs to specify the limits of integration in  \actconf.
For the hypersurface which asymptotes to the cylinder at $z=0$ defined by
$x_1^2+x_2^2<R^2$, the lower
limit is $r=0$ ( we will elaborate on  this in more detail below).
Near the boundary the solution of interest behaves like $z(r)\ra 0$ as $r\ra R$ and the integral
in \actconf\ diverges near $r=R$ and needs to be regularized.
We will introduce the upper limit of integration $r_a$ by
requiring that $z(r_a)=a$, where  $a\ra 0$ stands for short distance cutoff.
In general, entanglement entropy is sensitive to the short-distant modes
localized near the border of the region, and hence is cutoff-dependent.
We will need to understand the
behavior of $z(r)$ as $z\ra 0$ (and $r\ra R$).
The leading term is easy to get from \eomconf,
\eqn\zlead{ z\simeq 2\sqrt{R} \sqrt{R-r}   }
To go further in the expansion, it is convenient to introduce $x=1-R/r$
and make a substitution \eqn\subz{  z= 2 R \sqrt{x} f(x)    }
which brings \eomconf\ into the form
\eqn\eomconff{
\eqalign{ &4 x [x (1-x) f''(x){-}4 x^2 f'(x)^3{+}(4{-}5x) f'(x)] f(x) {+}f(x)^2 [2{-}4x {-}24 x^2 f'(x)^2]   \cr
              &\qquad   +3 (1{-}x) x (1{+}4 x f'(x)^2){-}12 x f(x)^3 f'(x) -2 f(x)^4=0     \cr}
}
\midinsert\bigskip{\vbox{{\epsfxsize=3in
        \nobreak
    \centerline{\epsfbox{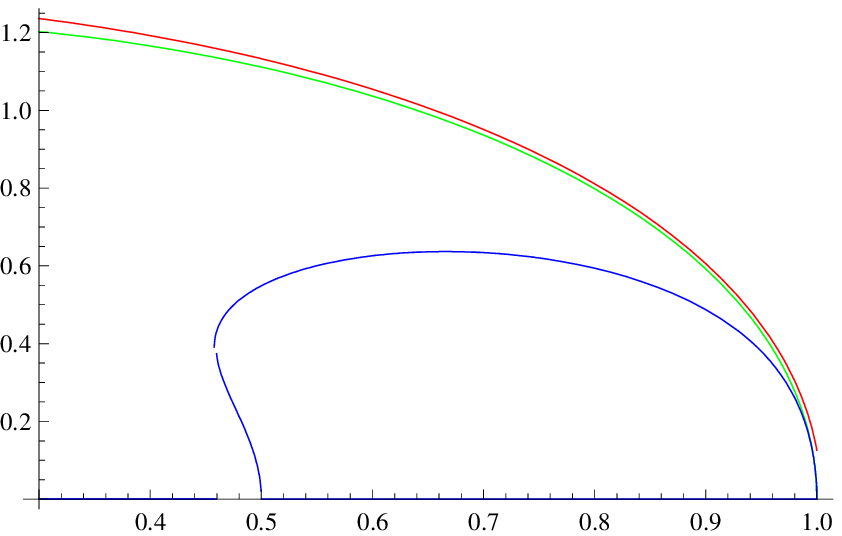}}
        \nobreak\bigskip
    {\raggedright\it \vbox{
{\bf Fig 1.}
{\it  z(r) for $R=0.5,\,\xi=7\times10^{-7}$ and $R=1,\,\xi=-6.1\times10^{-6}$ [blue curve];
            $R=1,\,\xi=-2\times10^{-6}$ [green curve] and $z(r=0)=1.29,z'(r=0)=0$ [red curve].
}}}}}}
\bigskip\endinsert
\noindent
This equation has a solution in terms of the expansion
\eqn\expsc{   f(x)= 1+ a_1 x +a_2 x \log(x) + b_1 x^2 + b_2 x^2 \log(x)+ b_3 x^2 \log(x)^2 +\ldots  }
where $a_2=1/8$ and other coefficients determined by the value of $a_1$.
Varying the value of $a_1$ we can vary the trajectory of the minimal
surface in the AdS space.
In practice, we specify initial conditions at the cutoff $z=a=10^{-3}$
in the form
\eqn\zin{  z(R)=a;\qquad z'(R)=-{2 R\over a} \left(1+\xi\right)  }
Varying $\xi$ effectively varies the trajectory $z(r)$ originating at $z(R)=0$.
A generic trajectory looks like the blue curve in Fig. 1.
It reaches a maximal value for $z$, then turns down and goes back to the boundary.
The corresponding hypersurface asymptotes to two cylinders
in the boundary theory.
Some particular value of $\xi$ corresponds to the surface which asymptotes to
the cylinder of radius $R$ at the boundary (this is the red curve in Fig.1, it can
also be obtained by requiring $z'(r=0)=0$).
There are, of course, curves with smaller values of $\xi$.
They look like the blue curve in Fig. 2, with the rescaled value
of the radius, $r(z)\ra 2 r(z)$, and asymptote to two cylinders on the boundary.
%
\midinsert\bigskip{\vbox{{\epsfxsize=3in
        \nobreak
    \centerline{\epsfbox{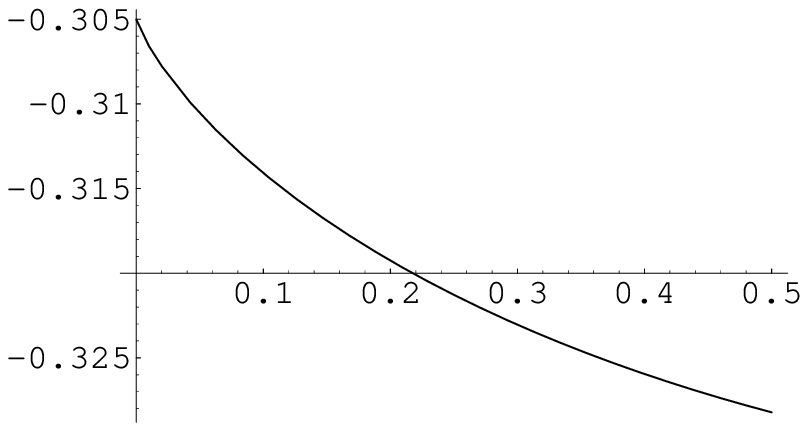}}
        \nobreak\bigskip
    {\raggedright\it \vbox{
{\bf Fig 2.}
{\it  $-\tilde\gamma$ as a function of $R-r_a$ for $R=1$
}}}}}}
\bigskip\endinsert
\noindent
In the following we will be interested in computing the value of entanglement
entropy for the solution that asymptotes to the cylinder.
It is a simple exercise  to extract the UV-divergent term (see also \SolodukhinDH).
For this purpose the parametrization $r(z)$ is more convenient.
The value of expression \entactionc\ in this parametrization is
\eqn\rzpara{  S= {4 N_c^2 l\over 15\pi} \int_a dz {r(z) \over z^3} \sqrt{ 1+{\dot r}^2 }}
where dot denotes derivative with respect to $z$ and $a$ is the UV cutoff.
Inverting \subz\ and \expsc\ we obtain
\eqn\expr{  r=R \left[ 1-{z^2\over 4 R^2}+\OO(z^4\log z) \right]  }
this expression can be substituted into \rzpara\ to obtain
\eqn\sdiv{  S= {2 N_c^2 \over 15\pi} \left({l R\over a^2}+ {l\over 4 R}\log {a\over R}\right) -{4 N_c^2 l\over 15\pi R} \tilde\gamma  }
where $\tilde\gamma$ is finite in the limit when short distance cutoff is
taken to zero, $a\ra 0$.
Numerical evaluation of the integral \rzpara\ and subtraction of
divergent terms in \sdiv\ produces a finite value of $\tilde\gamma=0.305$.
In  Fig. 2 we plot the (normalized) value of the integral in
\actconf\ and observe that it is indeed finite in the limit $a\ra0$, $r_a\ra R$.
It is interesting that the limiting value of $\tilde\gamma$ is positive and non-zero, similarly
to the topological entropy in \entent.
Of course, in the conformal theory the correlation length is infinite, so
it is not completely clear what the physical interpretation of nonvanishing
$\tilde\gamma$ is.

%
\subsec{Finite compactification radius}
\noindent Consider now the case with finite Kaluza-Klein radius $R_3$,
and hence finite correlation length.
For the
rest of this Section it will be convenient to pass to  rescaled variables,
\eqn\rescale{  r\ra {r\over z_0};\qquad z\ra {z\over z_0};\qquad R\ra {R\over z_0}  }
It is useful to remember that $z_0=2 R_3$ defines the scale of the correlation
length in the theory.
The rescaled coordinate $z$  is now bounded from above by $z=1$,
where the Kaluza-Klein circle shrinks to zero size.
Eq. \entactionc\ becomes
\eqn\acta{  S= {4 N_c^2\over 15} \int_0^{R\over z_0} dr {r\over z^3} \sqrt{ 1-z^4 +(z')^2 }}
The equation of motion which follows from \acta,
\eqn\eoma{  {d\over dr} \left( r z'\over z^3\sqrt{1-z^4+(z')^2}\right) ={r (z^4-3 [1+(z')^2])\over z^4\sqrt{1-z^4+(z')^2}}}
in general has to be solved numerically.
A detailed study of possible solutions to \eoma\ is relegated to the
Appendix.
Here we briefly summarize the results.
There are two types of solutions which asymptote to the circle of
radius $R$ in 2+1 dimensional boundary theory (they are shown in Fig. 3).
One type involves a connected surface which approaches the
circle of radius $R$ near the boundary $z=0$ and  has the disk topology
in the $(r,\phi)$ coordinates.
It corresponds to the blue curve in Fig. 3.
This solution can be found be starting at $r=0$ with boundary
conditions $z_*\equiv z(r=0)<1$, $z'(0)=0$ and
integrating \eoma\ outwards in $r$.
In Fig. 4 the value of $R$ is plotted as a function of $z_*$.
Similar to the slab geometry, studied in \KlebanovWS, there
appears to be a maximal value of $R$, beyond which this
solution does not exist.
Moreover, in the first approximation there are two branches for
sufficiently large values of $R$.
The new feature in comparison with \KlebanovWS\ is the existense
of the lower bound on $R$ for the branch which consists of the solutions
passing close to the end of the space at $z=1$.
We will call the solution with $z_*\ra 1$ critical, since, as we
will see below, it joins with the  solution which has the
topology of the cylinder. (see below).
We denote the value of $R$ for such critical solution by $R_c$.

Another interesting solution, which was absent in the conformal case
involves the surface which starts at $z=1, r=r_0$ and goes all the way
to the boundary.
This surface has the topology of the cylinder in the $(r,\phi)$ coordinates
and is an analog of the disconnected solution in \KlebanovWS.
It corresponds to the red curve in Fig. 3.
To exhibit this solution (and for the purpose of
computing the entanglement entropy below) it is convenient to use parametrization $r(z)$,
like in \rzpara.
In the non-conformal case the action is given by
\eqn\ncrzpara{  S= {4 N_c^2 \over 15\pi} \int_a dz {r(z) \over z^3} \sqrt{ 1+(1-z^4) {\dot r}^2 }}
and the equation of motion
\eqn\eomrznc{  {d\over d z}  {(1-z^4) r {\dot r}\over z^3 \sqrt{ 1+(1-z^4) {\dot r}^2 }}=
        {\sqrt{ 1+(1-z^4) {\dot r}^2 }\over z^3}  }
Near $z=1$ \eomrznc\ has a solution
\eqn\rshor{  r=r_0+{1-z\over 4 r_0}+\ldots }
where $r_0$ is the value of $r$ at $z=1$ and the dots stand for the terms subleading in $1-z$.
Starting with these boundary conditions and integrating \eomrznc\ or
\eoma\ numerically we obtain the solution with the topology of the cylinder.
Note that at large $r$ the solution of \rshor\ can be found in the
form
\eqn\lrform{  r= r_0 + {f(z)\over r_0} + o(1/r_0)  }
We can also verify \lrform\ numerically.
In Fig. 5 we show the behavior of $R$ as a function of $r_0$.
At large $r_0$ we recover \lrform, but at small $r_0$ there is
some interesting structure.
As $r_0\ra0$, the value of $R$ approaches $R_c$ and the solution
approaches the critical solution (with the disk topology) discussed earlier.
As $r_0$ increases, $R$ goes down, before climbing up back again.

In addition to the two solutions described below,
and just as in the conformal case studied above, there are also solutions which
approach the two concentric circles on the boundary.
Such solutions will define an annulus in the boundary 2+1 dimensional
theory.
We will not discuss them further; more details are provided in the appendix.
\noindent
\midinsert\bigskip{\vbox{{\epsfxsize=3in
        \nobreak
    \centerline{\epsfbox{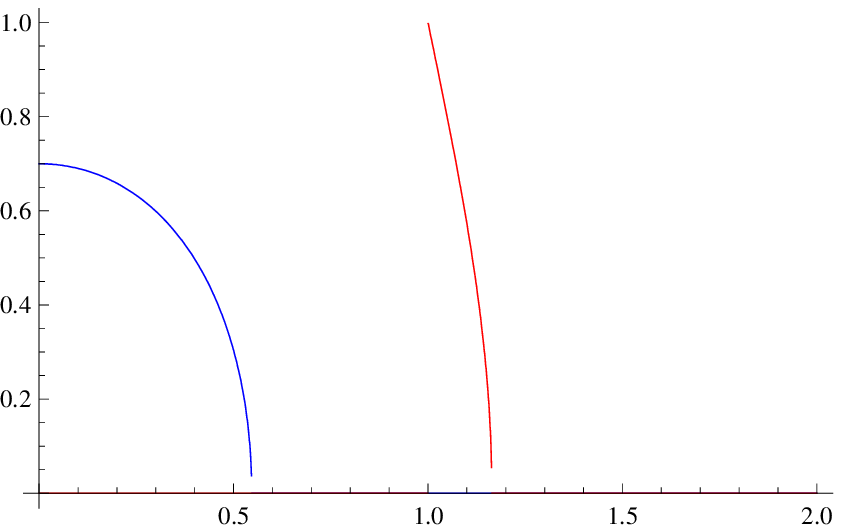}}
        \nobreak\bigskip
    {\raggedright\it \vbox{
{\bf Fig 3.}
{\it  $z(r)$. Disk [blue] and cylinder [red] topology.
}}}}}}
\bigskip\endinsert
\noindent
\midinsert\bigskip{\vbox{{\epsfxsize=3in
        \nobreak
    \centerline{\epsfbox{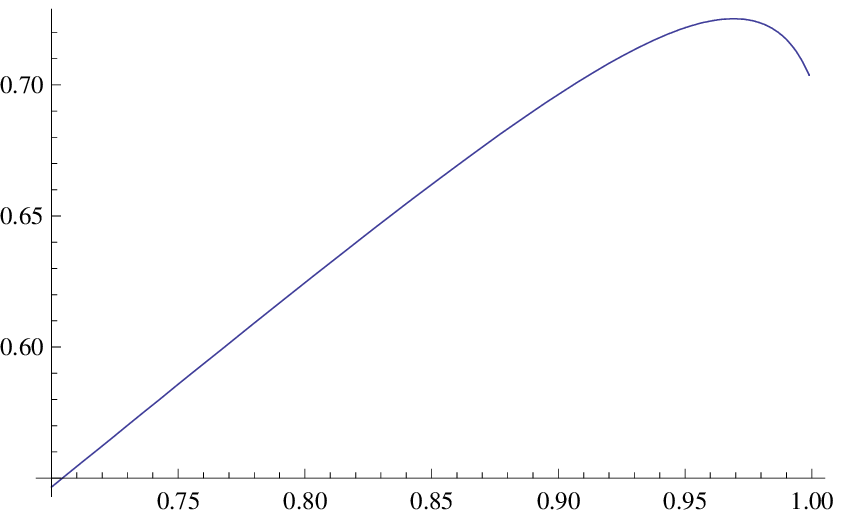}}
        \nobreak\bigskip
    {\raggedright\it \vbox{
{\bf Fig 4.}
{\it  $R$ as a function of $z_*$ (solutions of disk topology)
}}}}}}
\bigskip\endinsert
\noindent
\midinsert\bigskip{\vbox{{\epsfxsize=3in
        \nobreak
    \centerline{\epsfbox{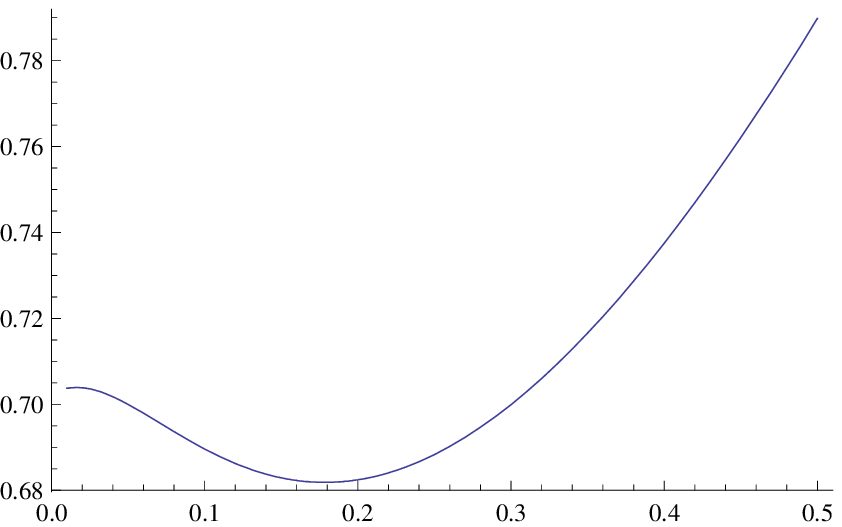}}
        \nobreak\bigskip
    {\raggedright\it \vbox{
{\bf Fig 5.}
{\it  $R$ as a function of $r_0$; (solutions of cylinder topology)
}}}}}}
\bigskip\endinsert
\noindent
\midinsert\bigskip{\vbox{{\epsfxsize=3in
        \nobreak
    \centerline{\epsfbox{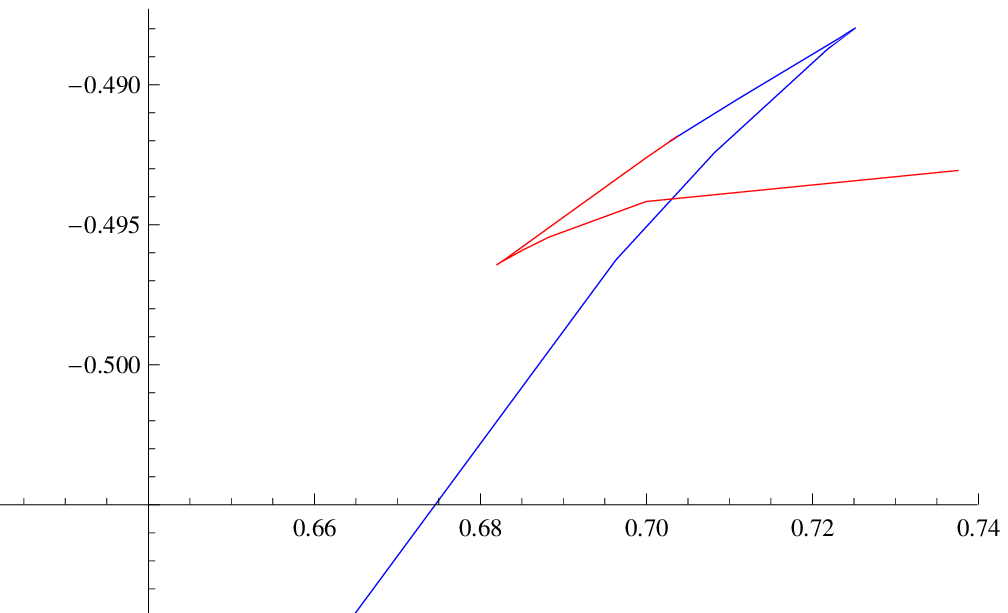}}
        \nobreak\bigskip
    {\raggedright\it \vbox{
{\bf Fig 6.}
{\it  ${\tilde S}$ as a function of $R$;
}}}}}}
\bigskip\endinsert
\noindent
%
%
To compute the entanglement entropy we follow the steps of
the previous subsection, where the conformal case was discussed.
In particular, the divergent terms are again given by~\sdiv.
We define the normalized finite part of the entropy $\tilde S$ via
\eqn\defts{  S= {4 N_c^2 \over 15 } \left({R R_3\over a^2}+ {R_3\over 4 R}\log {a\over R}\right) +{4 N_c^2 \over 15 } {\tilde S}  }

The value of ${\tilde S}$ as a function of $R$
[normalized according to \rescale] is plotted
in Fig. 6.
The blue part of the curve corresponds to the solution which
has the disk topology [blue curve in Fig. 3].
It is (modulo some fine details) double-valued, since $R$ is a double-valued
function of $z_*$, as can be inferred from Fig. 4.
For a given value of $R$ the branch with smaller $z_*$ dominates the
computation of entanglement entropy.
The solution with $z_*\ra1$ smoothly connects to the solution which has
the topology of the cylinder (red curve in Figs. 3,6).
The value of $\tilde S$ for this solution is again double-valued, according
to Fig. 5.
As the value of $R$ is increased, the cylinder solutions begins to dominate
the entanglement entropy (this happens at the point where the red line crosses the blue in Fig. 6).
For larger values of $R$ only the solution with cylinder topology exists.
According to \lrform, for large $R$,
\eqn\lrforma{   r=R+\OO(R^{-1})   }
and hence
\eqn\tildes{  {\tilde S}= -{R\over2}+\OO(R^{-1})    }
which implies vanishing topological entropy $\gamma$.

\newsec{D4 branes on a circle}
In this section we will consider the entanglement entropy of a three-dimensional ball in a confining theory in $d=3+1$.
The theory is obtained from compactifying a $4+1$ theory with radius $R_4$ \WittenZW.
At strong coupling, it has a gravitational description with metric,
\eqn\metricb{
\eqalign{
ds^2 &= \left({ U\over R}\right)^{3/2} \left[ \left({R\over
U}\right)^3 {dU^2\over f(U)}+dx_\mu dx^\mu+f(U) dx_4^2\right]
                   +R^{3/2} U^{1/2} d\Omega_4^2,\qquad \mu=0,1,2,3   \cr
e^{-2\phi} &=  \left({R\over U}\right)^{3/2}
}}
where $x_4 \sim x_4 + 2 \pi R_4$ and
\eqn\fdef{  f(U) = 1-\left({U_0\over U}\right)^3,\qquad U_0 = {4 \pi\over 9}{\lambda  \over R^2_4 }  }
The constant R is given by
\eqn\Rdef{ R^3 = \pi \lambda, }
where $$\lambda \equiv (2 \pi)^{-2}N_c g^2_{YM_5}= g_s N_c l_s$$ is the five-dimensional 't Hooft coupling.
The supergravity description is valid as long as~$e^{\phi} \ll 1$,
\eqn\ubound{
U \ll R \,.
}

The boundary is at $U \rightarrow \infty$, and
it is convenient to change the $(U,x^{\mu})$ coordinates into ~$(z,\tilde{x}^{\mu})$
defined by
\eqn\rescale{
\eqalign{
U & = \frac{U_0}{z}  \cr
x^{\mu}  & = \left(\frac{R^3}{U_0} \right)^{1/2} \tilde{x}^{\mu}
}}
We have now $z \in [0,1]$ and the boundary is at $z=0$. The metric and dilaton are now
\eqn\metricbres{
\eqalign{
ds^2 & = (R^3 U_0)^{1/2}  \left\{ z^{-3/2} \left[ {dz^2\over z(1-z^3)}+d\tilde{x}_\mu d\tilde{x}^\mu + (1-z^{3}) d\tilde{x}_4^2\right]
                   + z^{-1/2} d\Omega_4^2 \right\}
\cr
e^{-2\phi} &=  \left(\frac{R}{U_0} \right)^{3/2}z^{3/2}
}}

We consider dividing the $3$ spatial directions into a three-dimensional ball and its exterior, such that the surface between
the two domains is a two-sphere. Therefore, we are interested in an 8-dimensional surface in the bulk,
which wraps the inner 4-sphere, is constant along the periodic direction $\tilde{x}^4$ and ends in a two-sphere
of radius $\rho$ in the boundary $z=0$.

Let us call $r$ the radial coordinate in the $(\tilde{x}^1,\tilde{x}^2,\tilde{x}^3)$ directions.
The 8-dim surface is specified by the curve $z(r)$, or alternatively, by $r(z)$.
Choosing $z$ as the independent variable, the induced  metric is
\eqn\metricsphz{
ds^2  = (R^3 U_0)^{1/2}  \left\{ z^{-3/2} \left[ \left( {1\over z(1-z^3)} + \dot{r}^2(z)\right)dz^2 + r^2(z) d\Omega_2^2
+ (1-z^{3}) d\tilde{x}_4^2 \right]          + z^{-1/2} d\Omega_4^2 \right\}
}
and the area functional \entaction\ is
\eqn\actionsphz{
\eqalign{
S &= {1 \over 4 G_N^{(10)} } \int d^8\sigma e^{-2 \phi} \sqrt{G_{ind}^{(8)}} \cr
& = \frac{\pi^3 R^{6} U_0 R_4 }{2 G_N^{(10)}} \int \,dz \,   \frac{r^2(z)}{z^{3}} \sqrt{1 + z(1-z^3)r'^2(z) }
}}
where we have used $Vol(S^2)=\frac{4 \pi}{3}$ and $Vol(S^4)= \frac{8\pi^2}{3}$.
The equation of motion from this action is
\eqn\eomsph{
\eqalign{
& 4-4 z \left(-1+z^3\right) r'[z]^2 \cr
& +r[z] \left(2 \left(2+z^3\right) r'[z]
 +z \left(5-7 z^3+2 z^6\right) r'[z]^3+2 z \left(-1+z^3\right) r''[z]\right) =0
}}
Alternatively, choosing $r$ as the independent variable, the induced metric and the area functional are
\eqn\metricsphr{
ds^2  = (R^3 U_0)^{1/2}  \left\{ z^{-3/2} \left[ \left( 1 + \frac{z'^2(r)}{z(1-z^3)}\right) dr^2
+ r^2 d\Omega_2^2 + (1-z^{3}) d\tilde{x}_4^2 \right]          + z^{-1/2} d\Omega_4^2 \right\}
}
\eqn\actionsphr{
\eqalign{
S =\frac{\pi^3 R^{6} U_0 R_4 }{2 G_N^{(10)}} \int dr \,    \frac{r^2}{z^{3}(r)} \sqrt{ z(r)(1-z^3(r)) + z'^2(r)}
}}
and the equations of motion are
\eqn\eomphr{
\eqalign{
& 2 r z[r]^7+2 r z[r]^3 z'[r]^2 + 4 z'[r]^2 \left(r+z'[r]\right) \cr
& + z[r] \left(4 z'[r]+r \left(5+2 z''[r]\right)\right)-z[r]^4 \left(4 z'[r]+r \left(7+2 z''[r]\right)\right)=0
}}

\subsec{The two topologies}

The shape of the surfaces we are looking for is determined by the functions $r(z)$ or~$z(r)$. As in the previous Section,
starting with $z=0$ at $r=\rho$, there are
two possible shapes for $z(r)$:

$1)$  the curve starts from $z=0$ at $r=\rho$ and reaches $z=1$ at some value $r=r_0$. We call these curves
of 'cylinder' topology.

$2)$ the curve starts from $z=0$ at $r=\rho$ and ends at $r=0$ with $z=z_0\leq 1$. To avoid a conical singularity, we should demand $z'(0)=0$.
We call these curves of 'disk' topology.

In practice we can obtain the curves numerically. For the disk
topology, the initial conditions of the numerical problem are
$z(0)=z_0, \, z'(0)=0$.\foot{One can also try to find numerically the 'disk' solutions by fixing
the boundary conditions at $z=0$ (instead of at $r=0$), but the
shape of the curve is very sensitive to the initial conditions and a
slightly bigger or smaller value for $z'(\rho)$ does not give a
solution ending at $r=0$ with $z'(0)=0$.}

For the cylinder topology, the numerical problem can be set to find either~$z(r)$
or~$r(z)$, and in both cases we can set the boundary conditions either at $z=0$ or at $z=1$.
Consider the latter case first. With the boundary condition $z(r_0) = 1$, the leading term in the solution of \eomphr\ is
\eqn\solzrh{ z(r) \simeq 1 - \frac32 r_0(r-r_0) }
One can check that inverting this function,
\eqn\solrzh{ r(z) \simeq r_0 + \frac{2}{3r_0}(1-z)  \,,}
gives  the leading term of the solution of \eomsph\ with b.c. $r(1)= r_0$.
From the expressions~\solzrh\ or~\solrzh\
one can read out the boundary values of $z'(r_0)$ or $r'(1)$ needed for the numerical solution.

Alternatively, we can consider the behavior of the solutions near $z=0$.
The leading term in the solution of eq.\eomphr\ near the boundary ($z\sim 0$),
with boundary condition $z(\rho)=0$,
is\foot{There is a second solution to \eomphr\ that behaves as $z(r) \simeq -\frac14(\rho-r)^2$, but
it is not physical since $z \in [0,1]$.}
\eqn\solzrb{
z(r) \simeq \rho(\rho-r) + \cdots
}
The inverse to the above function has the expansion
\eqn\solrzb{
r(z) \simeq \rho - \frac{z}{\rho} - \frac{3z^2}{4 \rho^3}+ \cdots
}
which are  the leading terms of the solution of \eomsph\ with b.c. $r(0)=\rho$.
One can verify that the series \solrzb\ contains terms
of the type $z^n \ln^m z$, starting with $z^4 \ln z$.
From the expressions \solzrb\ or \solrzb\
one can read the boundary values of $z'(\rho)$ or $r'(0)$ which can be used
in the numerical solutions.

In Figure 7 we show the profile of $z(r)$ for solutions of both topologies.
As $z_0$ approaches $1$, there is a smooth transition between the two topologies, since the 'disk' solution
with $z_0=1$ coincides with the 'cylinder' solution with $r_0=0$.

\midinsert\bigskip{\vbox{{\epsfxsize=3in
        \nobreak
    \centerline{\epsfbox{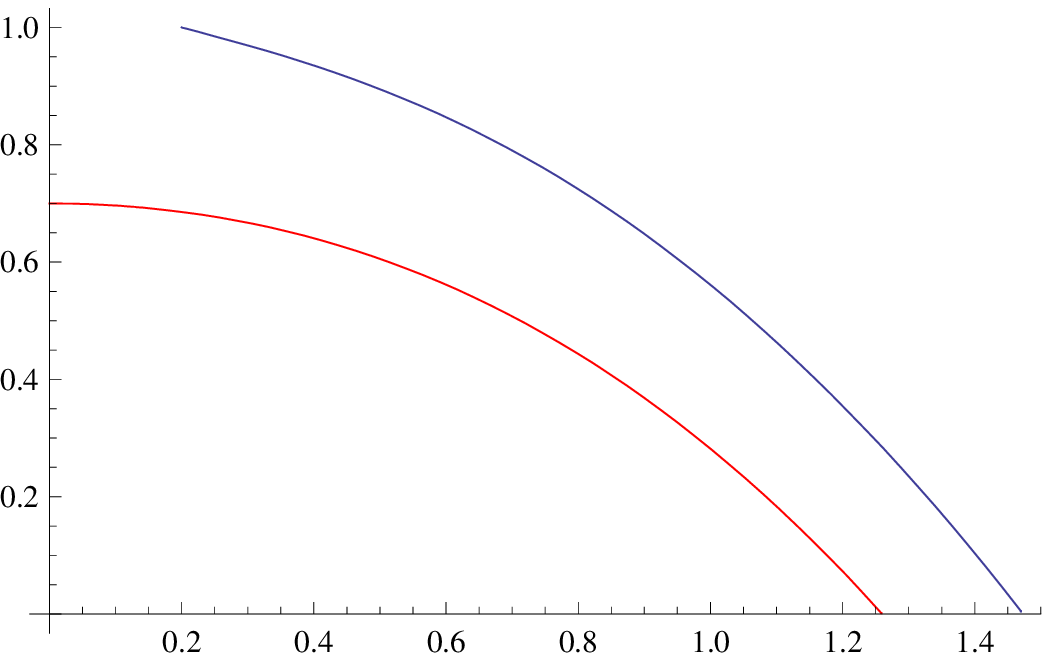}}
        \nobreak\bigskip
    {\raggedright\it \vbox{
{\bf Fig 7}
{\it Shape of $z(r)$ for a 'cylinder' solution with $r_0=0.2$ [blue] and for a 'disk' solution
with $z_0=0.7$ [red].
}}}}}}
\bigskip
\endinsert


The value of $r(0)=\rho$ has an interesting dependance on the value of $z_0$ and $r_0$ for the 'cylinder' and 'disk' solutions,
respectively, as we show in Figs. 8 and 9.
\midinsert\bigskip{\vbox{{\epsfxsize=3in
        \nobreak
    \centerline{\epsfbox{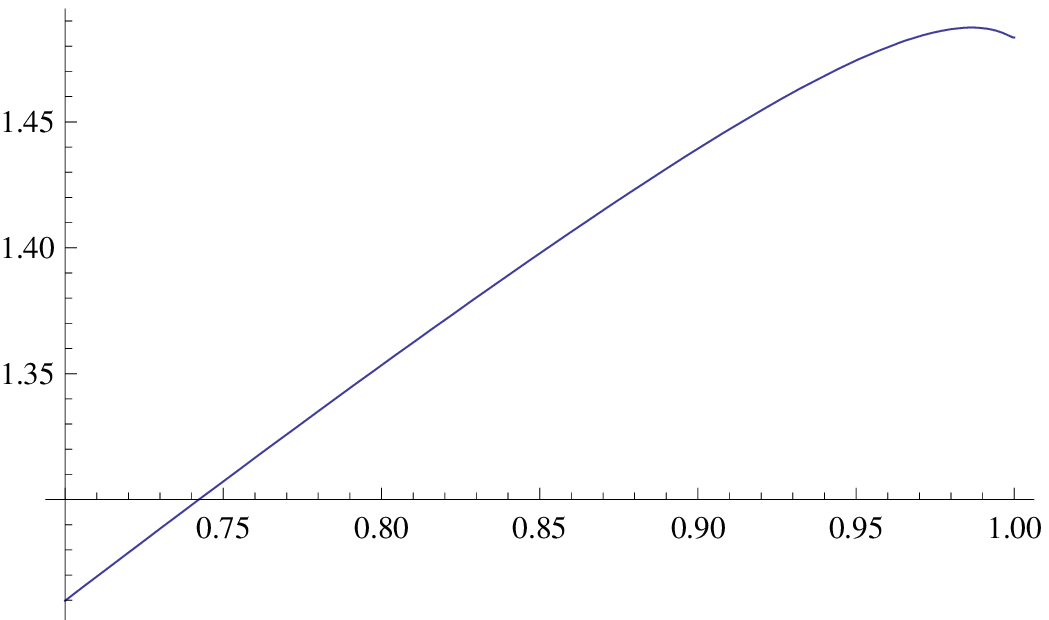}}
        \nobreak\bigskip
    {\raggedright\it \vbox{
{\bf Fig 8}
{\it  $\rho$ as function of $z_0$ for the 'disk' solutions.
}}}}}}
\bigskip
\endinsert
\midinsert\bigskip{\vbox{{\epsfxsize=3in
        \nobreak
    \centerline{\epsfbox{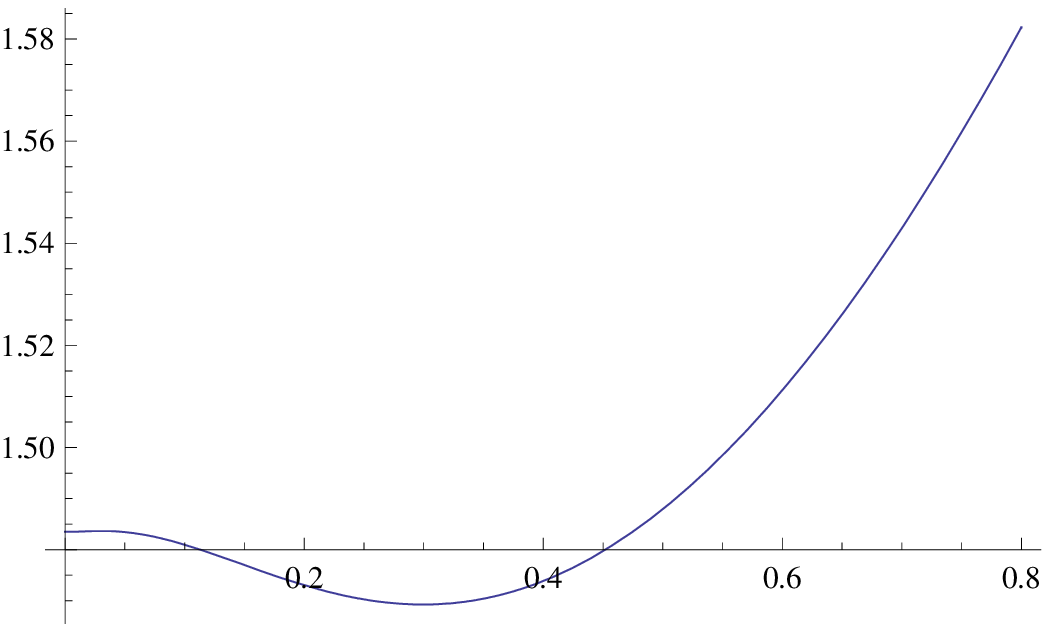}}
        \nobreak\bigskip
    {\raggedright\it \vbox{
{\bf Fig 9} $\rho$ as a function of $r_0$  for the 'cylinder' solutions.
{\it
}}}}}}
\bigskip
\endinsert
We see that for the 'disk' solutions  there is a maximal value $\rho$ as $z_0$ grows, and for bigger $z_0$ the value of $\rho$ shrinks.
A more detailed numerical study, which is not evident in Fig. 8, shows that the value of $\rho$ actually oscillates with decreasing amplitude
as~$z_0 \rightarrow 1$.
For the 'cylinder' solutions, $\rho$ grows monotonically with $r_0$ for large $r_0$. As $r_0$ diminishes, there is a minimum value
for $\rho$, such that for smaller values of $r_0$, $\rho$ starts to grow again. Also here a close look shows that the value
of $\rho$ oscillates with decreasing amplitude as~$r_0 \rightarrow 0$.

\subsec{The phase transition}

The entanglement entropy is  UV-divergent  and we will be interested in subtracting the divergencies
and obtaining a finite quantity.
The divergent terms in the entanglement entropy can be obtained by inserting the leading behavior \solrzb\
into \actionsphz. This gives
\eqn\entropysph{
\eqalign{
S & =\frac{\pi^3 R^{6} U_0 R_4 }{2 G_N^{(10)}}  \int_a^1 dz \, \frac{r^2(z)}{z^{3}} \sqrt{1 + z(1-z^3)r'^2(z) } \cr
& =
\frac{\pi^3 R^{6} U_0 R_4 }{2 G_N^{(10)}}  \left[ \frac{\rho^2}{2 a^2} - \frac{3}{2a} + \frac{ \ln a}{8 \rho^2} \right] + S_{finite}(a)
}}
where $S_{finite}(a)$ is finite as $a \rightarrow 0$.\foot{It is interesting
that from the $r^2$ and $r'^2$ dependence in \entropysph\ there are
contributions from the $z^2$ term in  \solrzb\ to the $\ln a$ term in \entropysph, but one can check that they cancel
and the divergence would be the same had we kept only the linear term in \solrzb.}
Unlike the slab geometry, the divergent term does depend on the size $\rho$. In Fig. 10 we show the
value $S_{finite}(0)$ as a function of $\rho$. Note that this is a multivalued function, since a given
 $\rho$ can  correspond to two different values of $z_0$ or $r_0$, as we saw above.

\midinsert\bigskip{\vbox{{\epsfxsize=3in
        \nobreak
    \centerline{\epsfbox{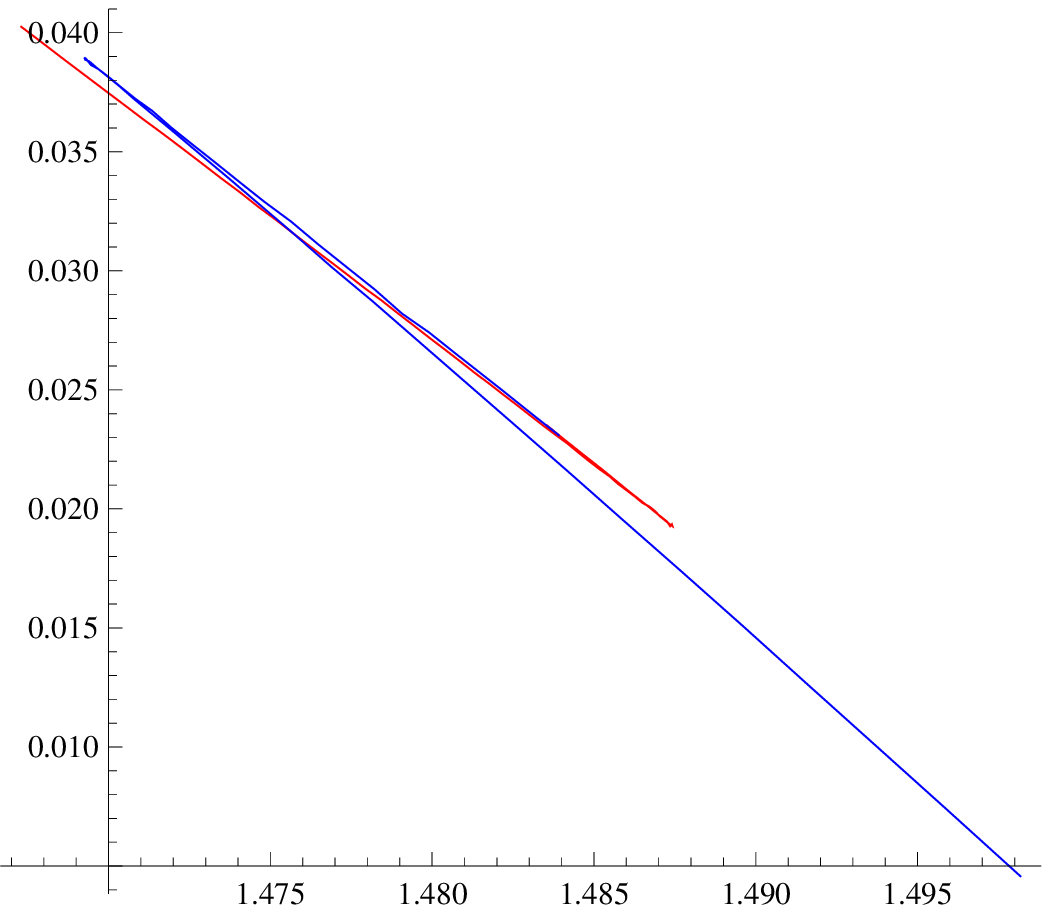}}
        \nobreak\bigskip
    {\raggedright\it \vbox{
{\bf Fig 10} The finite entanglement entropy $S_{finite}(0)$ as a function of $\rho$. The red line corresponds to the
'disk' topology and the blue line to the 'cylinder' topology.
{\it
}}}}}}
\bigskip
\endinsert

\subsec{Topological entropy}
To compute the topological entropy only the cylinder solutions are relevant, since
we are interested in the  the limit $\rho \rightarrow \infty$.
For the same reason, it is convenient to expand the solution in powers of $\rho$ as
\eqn\rhoexp{r(z) = \rho+ f(z) + \frac{g(z)}{\rho} + \frac{h(z)}{\rho^2} + \cdots
}
where $f(0)=g(0)=h(0)=\ldots=0$.
Inserting this expansion in the equation of motion \eomsph\ for $r(z)$, gives for the leading term~($ \sim \rho$),
\eqn\eqfp{
4 f'[z]+2 z^3 f'[z]+5 z f'[z]^3-7 z^4 f'[z]^3+2 z^7 f'[z]^3-2 z f''[z]+2 z^4 f''[z] =0
}
This equation is solved by $f'(z)=0$, and from $f(0)=0$ we get  that~$f(z)=0$.
With this result, the next term in the equation of motion \eomsph\ for $r(z)$ ($\sim \rho^0$) is
\eqn\eqfg{
4+4 g'[z]+2 z^3 g'[z]-2 z g''[z]+2 z^4 g''[z] =0
}
which has the exact solution
\eqn\solgp{
g'(z) = -\frac{1+c z^2}{1-z^3} \,,
}
and $c$ is an integration constant.
The 'cylinder' solutions are  required to be regular at~$z=1$, and this fixes $c=-1$, so we have
\eqn\solgp{
g'(z)= -\frac{1+z}{1+z+z^2}
}
Integrating this function gives
\eqn\solg{\eqalign{
g(z)& =-\frac{\arctan \left[\frac{1+2 z}{\sqrt{3}}\right]}{\sqrt{3}}-\frac{1}{2} \log\left[1+z+z^2\right] + \frac{\pi }{6 \sqrt{3}} \,, \cr
& = -(z+\frac{z^4}{4}+\frac{z^7}{7} + \cdots ) + (\frac{z^3}{3}+\frac{z^6}{6}+\frac{z^9}{9}+ \cdots )
}}
where the integration constant is fixed by $g(0)=0$.

As it turns out, the 'topological' term in the  entanglement entropy  depends only on~$g(z)$ and therefore
can be computed exactly.
Indeed, inserting the expansion~\rhoexp\ into~\actionsphz\ gives, using $f(0)=0$,
\eqn\entexp{\eqalign{
S & = \frac{\pi^3 R^{6} U_0 R_4 }{2 G_N^{(10)}}
 \int_a^1 dz \left[ \frac{\rho^2}{z^3}+\frac{4 g[z]+z \left(1-z^3\right) g'[z]^2}{2 z^3} \right] + O( \rho^{-1}) \cr
& = S_{\rho^2} -\gamma  + O( \rho^{-1})
}}
where
\eqn\sterms{\eqalign{
S_{\rho^2} & = \frac{\pi^3 R^{6} U_0  R_4 }{4 G_N^{(10)}} \rho^2 \left(  \frac{1}{a^2} - 1\right)
 = \frac{\pi^3 R^{3}  R_4   r_m^2}{4 G_N^{(10)}} \left( U_{\infty}^2 - U_0^2\right)
\cr
\gamma & = -\frac{\pi^3 R^{6} U_0 R_4 }{8 G_N^{(10)}} \left( \frac{\pi}{3} +2 + \log 3 -  \frac{6}{a} \right) \cr
& = - \frac{\pi^3 R^{6} U_0 R_4 }{8 G_N^{(10)}} \left( \frac{\pi}{3} +2 + \log 3 -  \frac{6 U_{\infty}}{U_0} \right)
}}
where $U_{\infty}= U_0/a$,
and the dimensionful radius of the ball is $r_m^2= \rho^2 R^3/ U_0$.
In order to express the entropy using gauge theory quantities we should use
\eqn\gnewton{ G_N^{(10)}= 8 \pi^6 g_s^2 = \frac{8 \pi^6 \lambda^2}{N_c^2}  }
and the relation between $U_{\infty}$ and the field theory cutoff $\Lambda$  in our background \PeetWN,
\eqn\eu{ U_{\infty} = g_{YM_5}^2 N_c \Lambda^2 = (2\pi)^2 \lambda \Lambda^2 }
This gives finally
\eqn\stermsf{\eqalign{
S_{\rho^2} & = \lambda N_c^2 r_m^2 R_4 \left( \frac{\pi^2 \Lambda^4}{2}  -\frac{1}{3^4 R^4_4} \right) \cr
\gamma & = - \frac{\lambda N_c^2 }{144 R_4} \left(\frac{\pi}{3} + 2 + \log 3 - 9 \pi \Lambda^2 R^2_4 \right)
}}
So we find a non-zero and cutoff dependent topological entanglement entropy. But we should remember that
the supergravity description can only be trusted in the limit \ubound, which for the cutoff $\Lambda$ means,
\eqn\cbound{
\Lambda \ll \frac{1}{(\lambda \alpha')^{1/3}} \,.
}

\newsec{Discussion}

In this paper we computed the entanglement entropy for a disk in 2+1 dimensions
and a ball in 3+1 dimensions in the confining theories with holographic
duals.
We observed a phase structure similar to the one found in \KlebanovWS,
with two types of solutions.
At small values of $R$ the solution of the disk topology dominates the
computation of entanglement entropy, while at large values of $R$ the cylinder
type solution dominates.
In 2+1 dimensions this structure ensures vanishing topological entropy,
which is an expected result for a QCD-type theory.
This can also be viewed as a consistency check of the holographic prescription.
(Of course, it would be great to make progress in understanding the
prescription of Takayanagi and Ryu from the first principles.)
It is interesting that in the conformal case the analog of the topological entropy
is nonzero.
It is not completely clear what the physical interpretation of this fact is.

In 3+1 dimensions one can define a quantity which is analogous to
topological entropy. It turns out to be UV divergent.
A possible explanation of this behavior is that, as mentioned, the supergravity description of the
field theory cannot be trusted beyond the bound \cbound.
On the other hand, the topological entropy has not been studied much in 3+1 dimensions,
and such studies might shed light on our result.

Both in the 2+1 and 3+1 cases the solution with the disk topology joins the solution with
the cylinder topology at some critical value of the disk or ball radius.
The behavior of the entanglement entropy near this point seems reminiscent of
self-similar type of behavior of free energy of probe branes, described in \MateosNU.
It would be interesting to investigate this in more detail.

For the slab geometry, signatures of the phase transition predicted in \KlebanovWS\ have recently been found
in numerical studies in the lattice \BuividovichKQ\ (see also \VelytskyRS). It would be interesting to seek for similar signatures
of the phase transition predicted by the disk and spherical geometries.
But it should be noted that, unlike the case of the slab geometry \KlebanovWS, in our case the (regularized)
entropy scales as $N_c^2$ at both sides of the critical point. This suggests that the existence
of the phase transition might be unrelated to the confining properties of the theory.
Studying entanglement entropy for other geometries, different from disks and slabs,  might also shed light on
the relation of entanglement entropy to the properties of the theory.

Finally, it would be interesting to find supergravity backgrounds which would
correspond to theories with  non-zero topological entanglement entropy. One instance could  be the system of D2-D6 branes
which models the Quantum Hall effect \BrodieYZ, but to apply the methods of this paper
one would need the full supergravity solution, and not only the probe approximation used in \BrodieYZ.

\bigskip
\bigskip

\noindent {\bf Acknowledgements:} We thank S. Giombi, C. Herzog, I. Klich, M. Levin, M. Kulaxizi and D. Kutasov for discussions.
Ari Pakman thanks Harvard University for hospitality. The work of Ari Pakman is supported by the Simons Foundation.
Andrei Parnachev thanks KITP, Santa Barbara, University of Amsterdam, Ecole Polytechnique and
University of Washington for hospitality during the course of this work.

\appendix\A{Solutions in detail}

\noindent
To investigate the structure of the solutions of \eoma\ and \eomrznc\  we repeat the analysis
of Section 3.1.
The UV behavior of the solution of \eoma\ still satisfies
\subz\ and \expsc, although eq. \eomconff\ is modified.
Typical behavior of the solutions for sufficiently small $R$ are shown
in Fig. 11.
The picture is completely similar to the one in Fig. 1, which is not
surprising, since $R$ is substantially smaller than the correlation length.
In Fig. 12 $r(z)$ is computed. In particular, there is a maximal
value of $z$ which the curves can attain.
The red and blue curves in Fig. 12 approximate to the annuli with
internal radius $R_i=R=0.468$ and external radius $R_e>R$.
The green curve in Fig. 12. (which is the counterpart of the green curve in Fig. 11)
gets very close to the connected solution but eventually approximates to the
annuli with $R_e=R=0.468$, $R_i\ll R$.
In Fig. 13 we plot the structure of the solutions for the critical value
of $R=R_c=0.7034$.
The blue curve is essentially the critical solution, while the red curve comes
very close to $z=z_0$ before turning up and ending back on the boundary again.
It is instructive to see how this picture changes for $R>R_c$ in Fig. 14.
We now start with $R=1>R_c$.
The red line gives rise to the annulus with $R_i<R_e=1$, while
the blue line corresponds to $R_e>R_i=1$ and comes very close to
$z=z_0$.
The green line separates the two domains; it is natural to
think that it ends at $z=z_0$ and gives rise to a surface with
annulus topology in the bulk (not to be confused with the
topology of the boundary which is a circle).
This is the analog of the disconnected solution in \KlebanovWS.
This solution can be exhibited in a more
direct way.
One can analyze the equation of motion near $z=1,r=r_0>0$
and integrate the solution from there.
This is done in Section 3.2.
\noindent
\midinsert\bigskip{\vbox{{\epsfxsize=3in
        \nobreak
    \centerline{\epsfbox{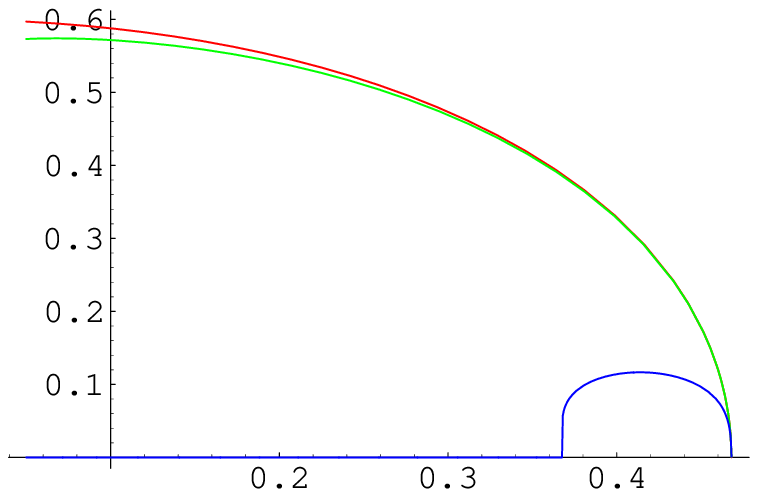}}
        \nobreak\bigskip
    {\raggedright\it \vbox{
{\bf Fig. 11}
{\it  $z(r)$ in confining background for $R=0.468$; $\xi=-5\times 10^{-4}$ [blue curve];
      $\xi=-8.05\times 10^{-6}$ [green curve]; $z'(r=0)=0$ [red curve]
}}}}}}
\bigskip\endinsert
\noindent
\midinsert\bigskip{\vbox{{\epsfxsize=3in
        \nobreak
    \centerline{\epsfbox{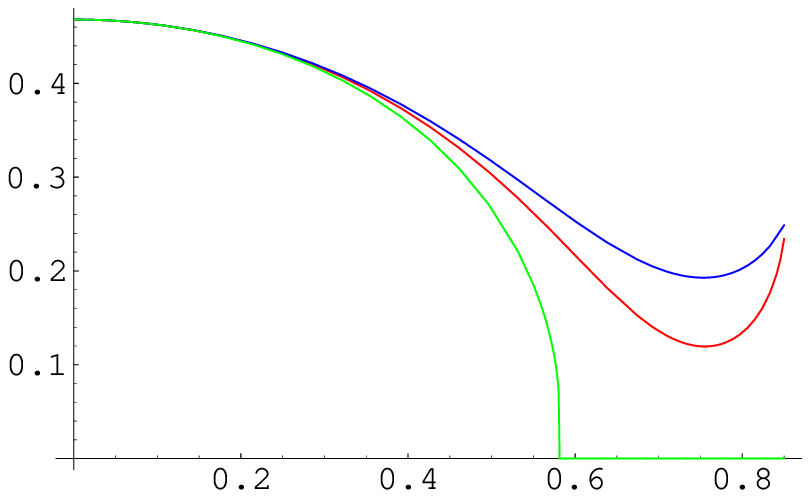}}
        \nobreak\bigskip
    {\raggedright\it \vbox{
{\bf Fig. 12}
{\it  $r(z)$ in confining background for $R=0.468$;
}}}}}}
\bigskip\endinsert
\noindent
\midinsert\bigskip{\vbox{{\epsfxsize=3in
        \nobreak
    \centerline{\epsfbox{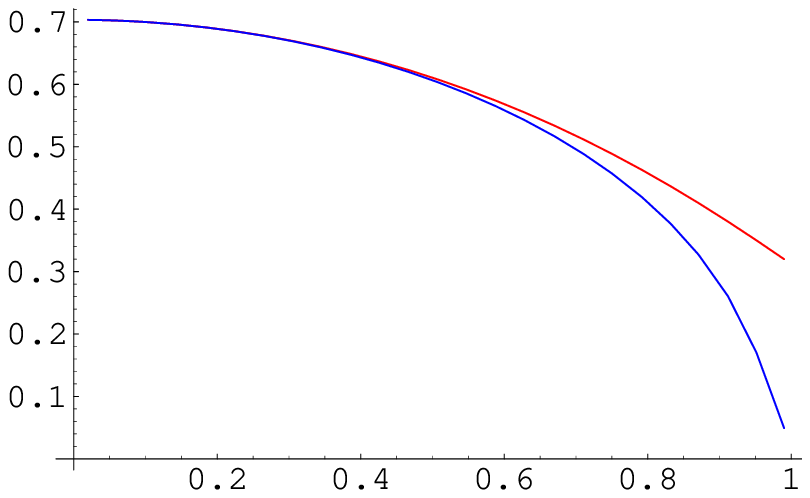}}
        \nobreak\bigskip
    {\raggedright\it \vbox{
{\bf Fig. 13}
{\it  $r(z)$ in confining background for $R=R_c=0.7034$;
}}}}}}
\bigskip\endinsert
\noindent
\midinsert\bigskip{\vbox{{\epsfxsize=3in
        \nobreak
    \centerline{\epsfbox{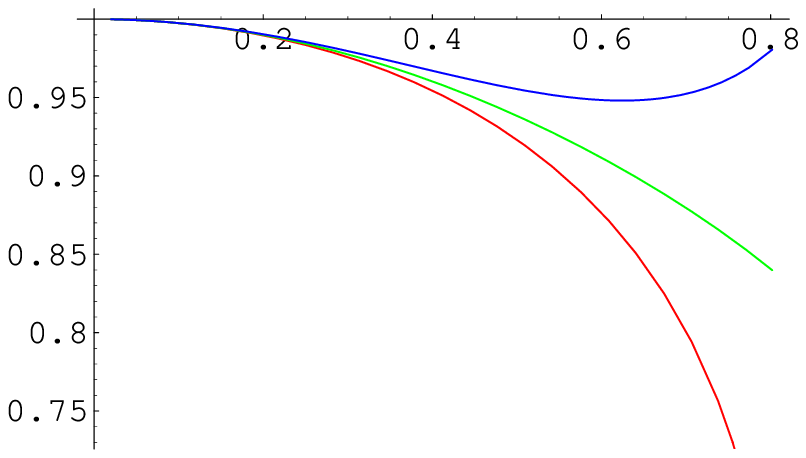}}
        \nobreak\bigskip
    {\raggedright\it \vbox{
{\bf Fig. 14}
{\it  $r(z)$ in confining background for $R=1>R_c$;
}}}}}}
\bigskip\endinsert

\footatend\vfill\supereject\immediate\closeout\rfile\writestoppt
\baselineskip=14pt\centerline{{\bf References}}\bigskip{\frenchspacing%
\parindent=20pt\escapechar=` \input refs.tmp\vfill\eject}\nonfrenchspacing
\end